\documentclass[12pt]{amsbook}

\usepackage{graphicx} 

 \numberwithin{equation}{chapter}
\numberwithin{figure}{chapter}

\begin{document}

\frontmatter 
\title[]{Virtual-Threading: Advanced General Purpose Processors Architecture }
\author{ Andrei I. Yafimau \newline Copyright \textcopyright  Andrei I.Yafimau 2009 }


\email{ eai\_inv@hotmail.co.uk }

\date{October 21, 2009}

\begin{abstract}

The paper describes the new computers architecture, the main features of which  
has been claimed in the Russian Federation patent 2312388 and in the US patent
application 11/991331. This architecture is intended to effective support of the 
General Purpose Parallel Computing (GPPC), the essence of which is extremely 
frequent switching of threads between states of activity and states of viewed in 
the paper the algorithmic latency. To emphasize the same impact of the 
architectural latency and the algorithmic latency upon GPPC, is introduced the
new notion of the generalized latency and is defined its quantitative measure - 
the Generalized Latency Tolerance (GLT). It is shown that a well suited for GPPC
implementation architecture should have high level of GLT and is described such
architecture, which is called the Virtual-Threaded Machine. This architecture
originates a processor virtualization in the direction of activities 
virtualization, which is orthogonal to the well-known direction of memory 
virtualization. The key elements of the architecture are 1) the distributed fine 
grain representation of the architectural register file, which elements are 
hardware swapped through levels of a microarchitectural memory, 2) the prioritized 
fine grain direct hardware multiprogramming, 3) the access controlled virtual 
addressing and 4) the hardware driven semaphores. The composition of these features lets to 
introduce new styles of operating system (OS) programming, which is free of 
interruptions, and of applied programming with a very rare using the OS services.


Subj-class: Architecture, Operating Systems 

ACM-class: C.1.2; D.4.1; D.4.2

\end{abstract}

\maketitle

\tableofcontents

\listoffigures

\mainmatter 

\newcounter{pnumb}                     	
\setcounter{pnumb}{\value{page}}       	
\mainmatter                            	
\setcounter{page}{\value{pnumb}}       	


\chapter{ Introduction }

Parallel computing was originally used in the mid-60s in specialized supercomputers that were 
primarily intended for massive parallel processing \cite{Enslow76}. However already in the beginning 
of the 90-s the well-known expert in the multithreaded architecture, Barton Smith, put forth the 
challenge about the necessity to create a supercomputer \cite{Smith90a}, \cite{Smith94}, which would be 
able to perform the general purpose parallel computation (GPPC) much more efficiently than the 
existing cluster systems based on universal microprocessors.  In his view, the emergence of such 
supercomputer should force a large influx of financial and human resources and should revitalize the 
industry's interest in a further development of a computer architecture. The challenge posed by 
Barton Smith has inspired the development of the innovative architecture presented in this article, 
which seems to be the architecture-candidate for the general purpose supercomputer 
implementation.

It should be clarified that in line with Smith \cite{Smith90a}, \cite{Smith94}, the GPPC refers  to the most 
widely used class of computations, represented by a large set of arbitrary and frequently switching 
processes and threads. At present these computations are most effectively supported in the multi-threaded architecture, which has been first applied in the mid-sixties to the implementation of the 
peripheral processors of supercomputer CDC-6600 \cite{CDC6600}. This architecture has been 
evolving  in the set of systems \cite{Smith82}, \cite{Smith88}, \cite{Tera90}, \cite{CrayMTA} and in the most advanced 
shape it has been implemented in the Cray Threadstorm \cite{CrayThr} microprocessor, which has 
been used in the special processing nodes of the heterogeneous Cray XMT supercomputer \cite{CrayXMT}.

Currently parallel processors are used widely - from supercomputers to household appliances. 
However all the processors while performing a real tasks, do not get high enough 
speed that a modern high-performance logic  should provide.  The well-known expert in the 
architecture of supercomputers, Thomas Sterling, in his paper \cite{Sterling05} identified the four major 
reasons explaining the low real efficiency of computers, which he calls 'the four horsemen of 
Apocalypse':

latency - delays  experienced while accessing memory or other parts of the system;

overhead - extra work due to management of concurrent operations;

contention - delays experienced due to competition over resources used on a shared basis;

starvation - hardware idle state because of insufficient parallelism as well as the lack of load 
balancing.

In the same paper \cite{Sterling05} he provides a number of solutions, whose implementation  should 
improve the actual performance of computers. This solutions include a hardware support of service activities 
(overheads), including the ISA for atomic compound operations on complex data, 
synchronization, communications and using of multithreading for latency suppression. Following this line, the present paper proposes the functionally 
complete system of hardware solutions, which should radically increase the efficiency of GPPC
implementations. 

The main difference between the general-purpose parallel computing and the massive parallel processing 
(MPP)  \cite{WikiMPP} is the significant, of an order of magnitude or more, exceedence of the number 
of the tasks capable of concurrent running over the number of processor's physical channels that can 
simultaneously carry out these tasks. Therefore, the decline in the productivity of even the simplest 
processors causes overheads associated with the extra work that has to be done to manage the 
program concurrency. In addition to the lack of hardware resources, an idle state of useful tasks 
can be caused by purely algorithmic reasons. These reasons include waiting by one program 
thread for an information from another thread about the ability to continue a work or about external 
events,  notified by means of IO devices. This article calls these states as states of algorithmic 
latency both of the works and the hardware, which is used to maintain these works in the latency state. 
The transit states related to switching the works between active and waiting 
states are also called the latency states. 

Rapid improvement of digital circuit engineering has resulted in great decrease of powerful 
microprocessors cost. The leading manufacturers of mass microprocessors Intel and AMD already 
produce multicore and multithreaded dies as their main products. Sun Microsystems introduced 
the 8-core UltraSPARC-T2 microprocessor  \cite{Niagara}, whose each core can perform 8 threads simultaneously. 
Accordingly, parallel processors have become very widely used, even in household electronics 
- laptops and smartphones. All these circumstances have led to very wide spread of parallel 
computing and force programmers to deal with parallel programming. In 
analyzing this phenomenon, in the key conference report ISC'07 Smith \cite{Smith07} put forth the
statement that the computer industry is faced with the challenge of the need to develop new 
concepts and architectures of parallel computing, which should greatly simplify programming 
of parallel systems of arbitrary size and complexity. This requirement to simplify significantly the 
GPPC programming together with the requirement to minimize the mentioned above four reasons 
of decline in performance, should be considered a generalized requirement for a new 
architecture.

On the basis of this generalized requirement we develop the innovative architecture of the 
processor (US patent pending), presented in this paper, which is called the Virtual-Threaded 
machine (VThM). The remainder of this article describes the basic concepts of this architecture, 
which should be considered as a significant improvement of multithreaded architecture in the 
direction of providing the tolerance to the algorithmic latency.  Main properties and functions of 
the architecture are  as follows:

1. The fine grain representation of architectural registers as a set of blocks in microarchitectural registers; this 
representation provides a distributed location of blocks on different levels of a microacrchitectural 
virtual memory; 

2. The inheritance of the thread-owner priority by all elements of the threads' representation 
in hardware, which representation includes the performing instructions and the fine grain representation of ISA registers;

3. The asynchronous hardware swapping of long time inactive elements of the distributed representation 
of the architectural registers on the levels of the virtual memory in accordance with their priorities and 
ageing metric; 

4. The fine grain processor decomposition, whose processing elements simultaneously interact with each other by 
means of transactions exchange via buffer pools of prioritized queues;

5. The hardware prioritized multiprogramming execution of a virtual set of threads in contexts of a 
virtual set of processes.

6. The prioritized fine grained switching of threads between states of activity and waiting with the accuracy closer to a few ISA instructions;

7. The full hardware support of an authorized access of running threads to a data of different processes.

8. The threads synchronization hardware features on the basis of the hardware-driven semaphores.

9. The hardware timing of the semaphore-related waiting states.

10. The hardware support threads data exchange with IO units and inter-processor communication without interruptions.

11. The hardware support for direct data exchange between processes' memory and IO devices, 
without using of drivers and an operation system; 

12. The uniform synchronization between  threads running in processors and threads running in IO devices.

The rest chapters of this paper contain the considerations, which enable us to formulate the 
concept and the features of VThM architecture. We also present an implementation of this  concept as a 
block diagram of major component's level. In conclusion we sum up the viewed decisions and define the
place of proposed architecture in the existing taxonomies.


\chapter{ Virtual-threading vs the "horsemen of Apocalypses" }

In existing architectures the main bulk of multiprogramming functions realizing the parallel 
execution of multiple tasks is implemented by means of software. The basic concept of the 
proposed architecture is that the all reasons of the degradation of the real performance considered above 
\cite{Sterling05} can be eliminated by means of the new hardware features which would support the deep 
virtualization of multiprogramming. The proposed features should bring the ratio of the software and 
hardware implemented functions of the multiprogramming at least up to the ratio achieved in the modern 
virtual memory management features. At the same time, efficiency of these features should be 
sufficient for the simultaneous execution of the very large, essentially the virtual, set of threads.

In other words, the proposed computer architecture should be balanced in the two orthogonal 
directions of virtualization - in the virtualization of memory management and in the virtualization of 
threads management. Since the threads management is almost completely transferred to the 
hardware level in the proposed architecture, we call this one the Virtual-Threaded Machine (VThM). 
Accordingly, we introduce the new term Virtual-Threading (VTh) to sum up the essence of the architecture functionality - the totally hardware-implemented multiprogramming. 

Let's consider briefly how the invention and development of the virtual memory hardware support 
features has simultaneously enhanced the efficiency of computers as well as simplified programming. On this basis, we describe the expected similar double effect resulting from the 
introduction of  virtual-threading - the increase in the real productivity of computers and the 
simplification of parallel programming. It seems that the most radical simplification of 
programming irrespectively of the level of algorithms encryption languages is associated with the 
first implementation of the concept of virtual memory in the Atlas computer, developed as early as  
1962 \cite{Kilburn62}. The simplicity of programming for machines with virtual memory consists in 
the  ability to write programs without worrying about whether the overall size of their code and 
data does not exceed the amount of a physical memory of a machine. The remaining necessity to 
ensure that this amount does not exceed the amount of the virtual address space size determined by 
the ISA, is not significant in most applications because of the very large size of this space. For 
example, a bubble-sorting program is very simple as long as  that the array is fully fit within a 
physical memory, but the programs' complexity increases by magnitude of several orders if the 
array does not fit within the physical memory and the machine does not support the virtual memory.

Since even the commodity and inexpensive modern microprocessors support the virtual addressing 
by the 64-bit address, a majority of modern day programmers almost ceased to care about the 
amount of required memory. However the simplicity of writing large programs brought by the virtual 
memory would be meaningless without the effective hardware support. The purely software operated 
overlay memory management \cite{Deitel90}, which was used before the invention of the virtual 
memory, greatly simplified the writing of large programs. But apart from the inconvenience of 
additional manual writing of scripts for overlay supervisor, this management was making 
significant overheads associated with the swapping of a large statically defined segments of memory. 
This mandatory correlation - simplification of parallel programming by means of the representation of 
a natural parallelism of an algorithm as a virtual set of threads and small overheads for 
organizing the concurrent run of this set - is the principal feature of the proposed VThM architecture.

The main reasons of the effectiveness of the modern hardware support of the virtual memory are the
multilevel organization one and the fine granulation of elements swapped between the memory levels. In the first 
computer with virtual memory \cite{Kilburn62}, the only swapping element was the page of about 3-kilobyte length. Apart from operating registers there was only one level of virtual memory - the 
main RAM, with which the central processor could interact at the microarchitectural level. The 
second level of the virtual memory was placed on the page drum and the operating system had to 
ensure page readings into the RAM by means of an ISA code sequence.

There are exist one to three levels of caches between the operating registers and RAM in modern 
computers, and swapped elements have the size of a few dozens bytes. The closer each level is 
situated to the processor, the smaller volume and the lower access time it has. In dynamics, the 
swapped elements are placed on the levels of the cache memory in such a way that the most 
frequently used items are placed on the upper fastest level and the least frequently used elements are 
pumped off into a RAM. The overall effect of using such a multi-level memory organization is the multiple increase in the speed of the program running due to the reduction a data exchange frequency
between operating registers and a RAM. It is important to note that in the proposed VThM architecture all movements between the 
levels of the cache memory, and between the lower-level caches and the RAM, are fully implemented by 
means of the hardware microarchitecture, and only the swapping of pages between the RAM and IO devices 
is realized by means of ISA code.

Following the brief description of the implementation of the virtual memory made above, it is 
convenient to clarify the concept of virtual-threading. In general terms, any works on a computer is 
represented by processes and threads, which essentially are the main virtual objects supported by an 
operating system. A set of all such works, which are represented by processes and threads, is 
known as a multiprogramming mix, whereas a computer system's work which executes the joint 
running of these works on a set of hardware units, is known as the multiprogramming. Prior to the
clarifications would be made in the following chapters, we assume that the multiprogramming mix 
consists of threads. 

Newly created threads and threads in the waiting state are represented by means of the data structures, 
hereinafter  referred to as the thread descriptors. These descriptors include the full file of architectural 
registers as well as a control information of the operating system. They are created by means of 
software and are located in a computer's multi-level virtual memory as a usual data.  To ensure 
that a  thread initially starts or resumes running after having been in the waiting state, the operating system 
must copy the architectural registers file from the descriptor in the memory into the hardware unit's physical 
registers file, thereafter referred to as the operating registers file. In the absence of a free operating 
registers file, the operating system should  copy one of the operating registers file occupied by a
descriptor of an another thread to the corresponding thread descriptor in the ISA level virtual 
memory and only after that the system can upload the other thread descriptor into the vacated 
place. This procedure of counter rewriting of registers is thereafter called the swapping of 
architectural registers.

The main drawback of existing architectures while performing the GPPC is the need of swapping of the
architectural registers as a whole. Indeed, with frequent transitions of a single thread between the 
active and waiting states, as a rule, in each state of activity  only a relatively small fragment of its code will 
be executed. Accordingly, this fragment will only work with a small subset of the architectural 
registers file, forming one or more blocks of the operating registers. For example, in the RISC 
architecture it could be the block of control registers, which contains the current command counter, 
the processor status word and the block of current window registers. In addition, this statement is 
confirmed by Annex H of the SPARC V9 architecture manual \cite{SPARCV9}, which shows the way 
of optimal code generation. In this code the leaf procedures can be made 
in such a way as to operate without their own register window, using their caller's window instead. 
Such leaf procedures will be the most frequently used fragments of optimized codes, which access only  
to the 1/8 of the ISA windows. 

It is clear that it possible to increase the efficiency of execution of the frequently switching threads 
code by means of fine grain organization of the operation registers file and corresponding 
organization of its swapping. This argument seems logical, because it is essentially the extension 
of the property of compactness of an active working set of virtual memory pages \cite{Dening68}, 
\cite{Dening80} to the fine grain representation of architectural registers files and microarchitectural 
operational registers files. Within such interpretation, the microarchitectural operation registers files 
constitute the fastest level of the virtual memory, and their blocks can be viewed as data cache blocks, 
which are dynamically loaded by the blocks of the corresponding architectural registers file by means 
of hardware  in accordance with executed instructions demands. 
Finaly, we can consider the VThM architecture as a result of an extremely deep  exploitation of the locality princip \cite{Dening05} in the processors microarchitecture implementation. 

In contrast to the optimized version of ther register file swapping as proposed above, in the existing 
architectures the entire file of architectural registers is always loaded into the operational registers 
file, regardless of the actual requirements of the fragment of code, which is being executed in the 
each phase of activity. In fact, such a coarse granularity of file registers swapping is fully consistent 
with the one-level virtual memory implementation in the first processors, in which the access to a 
small piece of data in an absent page forces loading of an entire page from the page drum into RAM 
\cite{Kilburn62}. This redundancy leads to time overheads for the swapping of a non-modified architectural 
register blocks and leads to increase of an idle hardware bulk, which is occupied by an unused blocks 
of the architectural registers.

The second drawback of the existing technique of multiprogramming is the program-implemented 
dispatching, in which the selection of a thread to be activated and a swapping of thread descriptors are 
performed by means of an ISA code just as the pages swapping was performed in the early virtual 
memory systems. This drawback is easily eliminated because the functions of the created threads  
dispatching, including the priority dispatching technique, are fairly simple. Their implementation 
by means of software uses only a small subset of ISA, which provides the integer arithmetic, load-store and synchronization instructions. This justifies the direct hardware implementation of the
dispatching functionality by means of  simple specialized unit, relieving the processor to execute 
the useful tasks. Within this approach, it is only the complex and rarely performed functions of 
operating systems that implement file systems, high level network protocols, resource accounting, 
as well as the creation and destruction of processes and threads, should be implemented by means 
of software.

Let us clarify the major aspects of the proposed fine grained threads dispatching. Creation of a thread 
leads to creation of the thread descriptor that contains the information necessary and sufficient for 
running the thread, with the possibility of suspending and resuming it purely by means of hardware. 
Such a descriptor  in fact is a virtual processor, stored in the multi-level microarchitectural virtual 
memory. An upper level of this memory is the fine grained operating registers blocks, into which the 
requested blocks of architectural registers are loaded by means of hardware in accordance with the 
dynamics of active threads. 

In fact, the virtual-threading virtualizes the fastest level of microarchitectural memory - the level 
of operating registers of the processor. As a result, the operating registers blocks, which store the 
fine grained threads descriptors elements, are swapped in the same manner as the data cache 
elements. Essentially this presentation is logically distributed - its elements can be allocated on 
different levels of microarchitectural virtual memory at each moment 
in accordance with the dynamics of threads' access to the architectural registers.

Summarizing what was said  above in this chapter, one can define the virtual-threading as the 
direct hardware management of threads from their inception to destruction, including their 
hardware dispatching and per-element hardware swapping of fine grain threads descriptors 
representation in accordance with requests of instructions streams. We suppose thay with a deep exploitation of 
so defined concept of virtual-threading in processor design, the de facto risen Silicon Curtain on both sides of which 
the hardware engineers and system programmers almost independently decide on the merits 
of the general problem of improving the efficiency of general purpose parallel computing,
should be eliminated. On one hand, the efforts of developers of processors led to the invention of the
multithreaded architecture (MTA) \cite{Tera90}, \cite{CrayXMT}. Such architecture provides the tolerance 
to the architectural latency, caused by long lasting accesses to the non-local memory in 
the NUMA-systems. For example, the Tera computer supports 128 simultaneous threads, which is 
enough to cover the maximum duration of access equal to 67 processor clock ticks. This 
architecture has broad scalability, which allowed to create the Cray XMT cluster system with few thousands 
nodes based on the MTA-processors \cite{CrayXMT}.

On the other hand,  joint efforts of developers of processors and system programmers led 
quite long time ago to the invention of multiprogramming, ensuring the tolerance to the algorithmic 
latency as defined above. However, this tolerance has been provided by means of 
software and hardware only at the macro level - at the level of large logical units, such as CPUs, IO
channels and IO units. This multiprogramming provides a parallel work of the potentially all 
units mentioned above by means of the hardware features, including  timers and IO interruptions as 
well as a direct memory access (DMA) channels. In fact, it provides a purely hardware concurrent 
work (CPU computations or IO units data transfers) for the benefit of threads, total number of 
which is not greater than the number of the hardware macro units. If the total number of the 
threads running in a system is greater than the number of these units, the current 
multiprogramming technique changes the running threads' set only by means of software. 

While performing the GPPC, the number of software threads exceeds the number of 
hardware threads by an order or more. Therefore the tolerance to the algorithmic latency can't be 
ensured by means of simple build-up of a number of hardware threads, as it is done in the MTA 
architecture to ensure the tolerance to the architectural latency. The common feature of the 
architectural and algorithmic latency is the consumption of computer resources in the latency 
states, e.g., in the states when the thread does not perform useful work. It is natural to call these 
states the generalized latency states. It should be noted that in a current MTA processors' implementations  resources, that provide 
tolerance to the architectural latency in a time interval, are busy in the interval and couldn't be 
simultaneously used to provide the tolerance to the algorithmic latency. 
	
For example, in the above mentioned Tera computer \cite{Tera90} the hardware thread can wait up to 64 
processor clocks while accessing to the non-local memory. During this time the hardware supporting 
the waiting state, is reserved and cannot be used to perform any ready-to-run thread. Therefore, 
the effective number of hardware threads, capable of providing tolerance to the algorithmic 
latency may be significantly smaller than it is defined by the MTA processor architecture.

Having thus defined the concept of generalized latency, it is naturally extend the general 
technical quality factor of any engines - the coefficient of useful work, E (Efficiency) - to 
computers architecture and determine it by the following simple formula:

E = 1 - Cl / C,

where C   - the total cost of hardware used in performing some quantity of works;

Cl  - the total cost of hardware remaining in a state of the generalized latency during performing this 
quantity of works.

The value of E can be considered the level of generalized latency tolerance (GLT), that is GLT = 
E. In the other words, one can suggest the following optimality principle of computer architecture, 
the most effective for performance GPPC. In the optimal architecture the set of works, defined as 
overheads, should be maximally reduced, while the remaining works, including keeping threads in the
generalized latency state, should be performed by a hardware of minimal cost.

In this context, the proposed VThM architecture can be defined as the architecture with high level 
of generalized latency tolerance. The key elements of the innovation, which induce the domino effect 
in the development of this architecture, are the distributed fine grained representation of the 
architectural register file, which is moved by means of hardware through the levels of the
microarchitectural virtual memory, and the direct hardware multiprogramming on the basis of this 
representation. The logical chain, providing the conceptual integrity of design, is as follows:

- the fine grain representation of the architectural register files and the direct hardware multiprogramming 
provide extremely fast and inexpensive switching of a threads between the activity and waiting states;

- the fast switching of threads provides the direct hardware support for a very large, essentially a virtual set 
of threads, which is sufficient for the direct representation of the total set of independent activities of 
parallel algorithms of any complexity;

- the effective hardware support of a virtual set of threads provides the uniform representation by means 
of threads for all activities of a different nature associated with the users' and system programs' 
threads, hardware interruption handlers, software signal handlers and hardware activities in  IO 
units;  

- the fast switching of threads provides the effective homogeneous synchronization of different nature 
threads by means of  the defined below hardware-driven semaphores;

- the uniform synchronization allows to abandon the concept of interruptions, and thus significantly to 
simplify programming, both that of operating systems, and of user applications.

The proposed architecture provides the balanced computer hardware virtualization in the two 
orthogonal directions - memory and activities. In the aspect of an implementation of practical 
processors it will cause the situation when the kernel algorithms of operating systems, like the 
algorithms of the multi-level virtual memory, will be built into the hardware. Accordingly, developers 
of the kernels of operating systems in addition to the use of the system programming languages 
assembler and c will use the hardware developers' languages such as verilog. This should bring 
together developers of software and hardware, and should move over time the mentioned above 
Silicon Curtain to the very appropriate place - in front of  "the four horsemen of Apocalypse". 

In the following four chapters the existing implementations of multiprogramming are analyzed and 
the key decisions that seem to be necessary for implementation in the VThM processors 
architecture are described.


\chapter{ Viability of virtual-threaded architecture}

After introducing the main conceptions we shall describe the main pragmatic and functional aspects which define the viability and possibility of a rapid spread of the virtual-threading in practical processors implementations.

We assume that one of the main pragmatic features of the proposed VThM architecture should be effective 
support of heterogeneous processors, where operating systems of native and legacy architectures 
may run simultaneously. Therefore this architecture should supports advanced features of 
virtual machines analogous to those used in processors \cite{Niagara}, with a purpose to supply an 
effective shared access control over physical resources at the hardware level. In 
architectures with virtual machines an overall management over all hardware resources is carried 
out by a special user, called thereafter the hyperuser. The resource management includes the creation of 
virtual machines, to which processors (or quants of processor time), memory and IO units are 
allocated. Managing the execution of programs, including that of operating systems, is done by the
virtual machine monitor \cite{VMM}, in such a way that these programs only have access to their allocated 
resources. 

Let us  define used in the VThM architecture the organization of the virtual machines management and 
organization of programs running on these virtual machines. In the conventional Unix clone operating systems, 
any work is performed on behalf of an authorized user, identified by the unique number - the User 
Identifier (UID). An operating system maintain accounts for all users and for all resources. It also provides 
a control over the users' access to the resources. The special user identifiable by UID=0, which is called 
a superuser or a system administrator, has an access to all resources. Such an operating systems in the  
VThM architecture are run on their own virtual machines. Accordingly,  superuser 
identifiers on each one has the local for this machine UID=0. The only one hyperuser has the global for all virtual machines UID=0. 

The execution of any work is always implemented by the three major, essentially virtual, entities - 
programs, processes and threads, each of which has a dedicated representation in the computer memory.  The term "virtual" reflects the fact 
that the entity has no a direct representation in the processor hardware, but the representation is created and destroyed 
by means of software in the memory. The work of a computer system providing for a simultaneous running is 
known as the multiprogramming mix management or shortly the multiprogramming. Let us consider  the 
basic concepts related to techniques of the hardware-software implementation of the multiprogramming 
in the existing computer architectures.

The program is the static object representing the result of translation of the source code written in 
programming languages into object modules with subsequent merging into the executable file - the
code file. The availability of the execution of a program by some user is controlled by the operating 
system. The single user task is implemented as a launching of the program's code file to run, which starts to 
execute an initial thread. The work of the thread, as of all created later ones, is the strictly sequential 
execution of the architectural instructions. In single-threaded processors such creation is done 
purely by means of software. In multithreaded processors there are special instructions, which 
provide the creation of new processes and threads \cite{Niagara}. All   threads that have been created 
in a process are executed in the context of this process. Each thread one can consider as the elementary  request of any work for using of 
processor or IO device resources. By means of using a plurality of threads-requests, one can speed up any work, whose 
algorithm allows a decomposition in  a form of concurrently running activities. 

A spawn of a new process results in the launching of its primary thread.  Running thread instructions 
can read and change the data in the context of thread's process-owner. This data is stored in the 
entities accessible on the ISA level - RAM and processor ore IO units registers. The current context of the available objects of the 
process is maintained under control of operating system. In fact the process is the elementary request 
of a user's task to work with a subset of computer objects accessible by means of software. The process 
protects user tasks  from uncontrolled mutual influences because of changing the data. At the 
same time, the processes can supply to each other parts of their own contexts with the 
purpose of sanctioned interaction. In the existing architectures  shared objects are implemented 
through the creation in each process, which uses the shared object, its own virtual address-alias 
affiliated with the common physical address of the shared object. Such object may be an area of non-resident memory, and as a result of the object swapping the memory management becomes significantly 
harder. This causes an increase in overheads associated with organization of multiprogramming. 

Let us clarify the terminology associated with the organization of the multiprogramming in the 
existing and proposed architectures. The concepts of the process and the thread are commonly 
accepted in majority of operating systems papers. But with the appearance of multicore and multithreaded 
processors, the terminology associated with support of their running by means of hardware units, 
varies significantly from one manufacturer to another. Below we define the relevant terms of 
process engine and thread engine. 

The process engine is defined as a collection of hardware features, which provide presentation of the 
separate process with ability to support the following features - a context which also called a protection domain and authorized controlled access to the context for other processes. 
This notion is close to the concepts of processor in the Tera architecture \cite{Tera90} and core in the architecture \cite{Niagara}.
 
The thread engine is defined as a collection of hardware features, which implement an execution of a 
separate software thread in the context of the process-owner and provide a necessary set of data types
and instructions types at the ISA level, sufficient to support correct and efficient synchronization. 
This notion is an extension of the concepts of hardware stream in the Tera architecture \cite{Tera90} 
and that of strand in the Niagara architecture \cite{Niagara} 

Using these two concepts of the process engine and the thread engine, the multiprogramming can be 
defined as athe concurrent  (or quasi concurrent) running of the set of virtual entities - processes and 
threads - by means of a hardware objects' set - process engines and thread engines. Let us 
consider the principal difference of the multiprogramming implementation in the existing and 
proposed architectures. In the existing processor architectures, the process engine and the tread engine 
are composed of functional logic units and at least of one file of the operational registers discussed 
above. In single-threaded processors all functional logic units work only with one operating 
registers file, which represents only one active process and one active thread. In multithreaded 
processors there is a plurality of operating registers file, which provide simultaneous running of the
corresponding plurality of active processes and threads without their program implemented 
swapping. For example, Tera MTA \cite{Tera90} and Cray Thread Storm \cite{CrayThr} provides 
concurrent running of 64 processes and 128 threads. 

The number of the operating registers files is the principal architectural limitation of the processor, 
which determines the static number of process engines and thread engines  and the corresponding number of a processes and threads capable to run simultaneously without program swapping. The operating 
registers files can be considered the highest and the fastest level of the microarchitectural memory of 
processor, whose special feature is the ability to represent and run active threads and processes. By  
contrast, in the proposed architecture the process engines and the tread engines are implemented in such a 
way that they independently support hardware swapping of the fine grained blocks of operation 
registers files. The immanent property of this swapping is its prioritized fine grained organization, such that registers file blocks associated with a thread of certain priority are unloaded into a lower level of the 
microarchitectural virtual memory if another thread with a higher priority requests the architectural 
register block when there is no free one. 

This swapping provides the deep virtualization of the process engine and thread engine and abolishes 
architectural restrictions on the number of hardware-supported processes and threads, which are 
immanent to all the existing architectures. In the proposed architecture, a plurality of hardware-supported processes and threads is determined only by resource constraints associated with the 
capacity of the executive pipeline to run the plurality of ones with a sufficient speed. 

Let us consider the existing techniques of a inter-process cooperation by means of the shared data areas and 
describe how in suggested architecture these techniques is moved on the hardware level. The 
process is identified by the unique number, has the attribute of affinity to the launching user 
and inherits the latter's access rights to computer resources. As pointed above, the 
value of UID=0 identifies the hyperuser, which manages all computer resources. All processes which runs as as proxy for the hyperuser can create the virtual machines and provide their with
dedicated resources. Each process has the following main features:

1. The context - a separate process is allocated its own context of available resources: its own virtual 
memory and objects identified by file names, the legal access to which is defined by the launching 
user; this user is identified by the attribute UID of process;

2. The direct protection - all processes are protected from uncontrolled mutual influence;

3. The authorized access - all processes can give each other the rights of access to the resources of their 
own context, e.g., can create shared objects;

4. The IO related indirect protection - a process can't initiate an IO which would lead to  
data changes in an other processes which are not announced to the process-initiator as shared. 

In accordance with the optimality criteria laid down in the previous chapter, in the proposed 
architecture only the rare actions of creating processes and descriptive information about their 
resource access sharing are implemented by means of software. However the special innovative 
hardware features are used for simultaneous storing of an information about all created processes and 
for simultaneous dynamic access control on the basis of this information. The cost of these 
innovative features is radically lower than  the cost of an entire processor, which is used in the 
existing combined hardware-software techniques for the same purposes. These features fully ensure 
the implementation of processors' features presented above (1-4). The similar using of the optimality criteria
is adopted in the hardware thread control described in the following chapter.  

The mechanism of hardware access control can be described as follows. A process as the main 
object of management for a process engine is characterized by the following attributes:

PID - the process identifier that uniquely identifies the process to the process engine;

PStat - the status of the process that determines the maximum status of the threads established in 
the process: hyper-privileged, privileged or non-privileged . Three status levels are required in 
order to support the virtual machines at the ISA level in the proposed architecture;

PPrior  - the process priority, which determines the maximum priority of threads created in it.

A thread as a primary unit of thread engine management is represented by the following attributes:

TID - the thread identifier, which uniquely identifies the thread for the thread engine. For clarity 
we assume that the most significant TID bits contain its process-owner identifier PID, and the 
remaining bits contain the local thread number TNo in the process-owner;

TStat - the status of thread, which - just like the status of a process - can be hyper-privileged, 
privileged and non-privileged, 

TPrior - the priority of a thread, which determines the order of  processing of the information 
related to the thread under a competition over any physical resources. 

The functions supported by the thread engine at the hardware level, include threads creation and 
completion, prioritized performing of calculations in all created threads, and organization of 
interaction between software threads as well as between software threads and IO threads. 

Launch of threads  is executed either automatically when the "Power" or "Reset" buttons are 
pressed, or by means of a program by  issuing relevant commands by a parent thread. In the 
first case, the so-called bootstrap threads are created, which have the hyperprivileged status. The 
hardware reset mechanism is implemented in such a way that the creation of bootstrap threads and the creation of specific bootstrap processes are performed as an atomic action. The context of 
this processes is the full context of physical resources of a machine available at the software level. 

The execution of threads at the ISA level boils down to their processing in two states - active and 
waiting, as well as their transfer between these two states. In the active state thread's commands 
are executed with the rate, which is determined by an overall number of threads in the active state as 
well as a number of hardware elements of thread engines capable of working in parallel. The 
emerging competition between several requests over the right to be served by elements of thread 
engines is organized by assigning the global priority to each atomic request and by using 
the buffer pool of queues at the entrance of each microarchitectural unit with independent activity. These queues are ordered 
according to the priority and the arrival of its elements. 

In contrast with such a fine grained representation, the existing mixed software-hardware thread's management uses  coarse grain representation, in which the priority of a thread is associated only with software visible thread descriptor. This leads to the emergence of a particularly unpleasant for real-time phenomenon known as priority inversion \cite{PriorInv}. 
Its essence is such that if a highest priority thread became ready to run 
when all thread engines are busy, it is necessary to perform swapping of the descriptor of this thread 
with a descriptor of another lower priority thread. In such cases, the operating system dispatcher is 
faced with a dilemma - either to perform swapping immediately or allow a lower priority thread to 
complete a time slice previously allocated by a scheduler and allow priority inversion during this 
time period. In the first case, overhead cost associated with switching at the expense of forced 
reduction of time slice, is increased. In the second case, the time of reaction decreases, which is 
crucially important for real time system and can forces its wrong functioning. The presented above fine grain thread representation and management automatically eliminates the priority inversion.


\chapter{ Interprocess communication implementation }

Having described the main objects of process and thread engines management, let us define how 
the information about these objects and about their access control can be compactly represented in 
the hardware.

The widespread Unix-cloned operating systems \cite{SunOS}, \cite{Linux} use file names to identify 
hardware computer devices visible at the ISA level, as well as to identify software-generated 
objects. A set of file names of the all objects can be uniquely mapped by an operating system to a set of a
numerical file identifiers FID, which will represent all objects in a computer system. Having 
made the FID as the part of the defined below access controlled virtual address, it is possible to ensure an
effective hardware dynamic control of access to all software and hardware objects of a computer.

The introduced below concept of access controlled virtual addressing is  the development of the 
structuring of virtual address space used in the Unix-like operating systems \cite{SunOS}, \cite{Linux}. In 
these operating systems the virtual spaces of the kernel and user processes are deferred only by a few 
most significant bits of a virtual address. 

An improvement, which is used in the VThM architecture, is that the process-destination of 
addressing is selected according to the number of the most significant bit of a thread-generated 
address, thereafter called the access controlled virtual address (ACVA).  If this bit, which we 
thereafter call VAShr, has a value of 0, the process-destination of addressing is the thread-owner 
process. This address is called thereafter the local ACVA and is represented by the two-component 
record

 (VAShr=0, LVA), where LVA is a virtual address in the process context. 

Correspondingly, the non-local address which is called thereafter the shared ACVA, is represented 
by the four-component record 

(VAShr =1, OPID, RefPID,  LVA), 

where OPID is the PID of the process-owner of the address, RefPID is the PID of the process-owner of 
the referencing thread, and LVA is the virtual address in OPID process context.

Using this mode of addressing, it is possible to consider that any shared object, which occupies L 
bytes of memory, can be represented on the hardware level by the following record 

(VAShr = 1, OPID, LVA, L). 

The dynamic access control to the shared objects can be implemented in such a way that the local address  ACVA
generated by any thread of RefPID process, which is identified by the zero value of VAShr, is 
controlled by the virtual memory management unit (VMMU) in the same way as in the existing 
architectures.

If the thread of the process with the identifier RefPID refers by the  shared ACVA, a reference 
implementing transaction is issued into the VMMU with the shared ACVA looks like 

(VAShr =1, OPID, RefPID,  LVA, LRef), where LRef is accessed data length.

The transaction includes the corresponding to the referencing instruction access code 
RefMode - read, write, execute or synchronization atomic access.

Such improved VMMU, thereafter called the memory and IO management unit (MIOMU), has the 
access control directory. This directory contains a set of access admitting records looks like 

(OPID, GntPID, OrVA, L, GntMode). 

Each such record defines that all threads of the process with the identifier GntPID 
have the ability to access the virtual memory area [OrVA, OrVA+L-1] of the process with the identifier OPID
with the instruction codes RefMode matching to the GntMode field. GntMode is the mask which
defines the subset of allowed referencing operation - read, write, execute or synchronization atomic access.
The transaction is performed if there is at least one appropriate record in the MIOMU. 

It should be noted that the processes (more precisely, threads), which have the hyper-privileged 
status (more precisely, threads with this status) can apply to all facilities accessible at the ISA 
level at their physical addresses without any control. The physical address can be presented as an 
address pair (ASI, PhA) similarly to the SPARC architecture \cite{SPARCV9} 

On the basis of the defined above ACVA in the VThM architecture an advanced technique of 
direct memory access (DMA) between any process memory and any IO devices is implemented. 
This technique supports the mentioned above indirect protection associated with input-output. 
Using the introduced above notations, we assume that the control registers block of any 
independent activity channel in any IO device can be represented by means of ACVA as a shared 
object (VAShr =1, OPID=0, LVA, L), the owner of which is the hyperuser represented in the
microarchitecture by the process identifier PID=0.  

If threads of a process will work with such a shared object, one of these threads should get an 
access to the object by using a corresponding software executable system call.  If an operating 
system permits this request, it will create the permitting record in the MIOMU and will return the 
shared ACVA of the IO unit channel control registers block. Similarly, it is possible to grant 
access to the same IO unit for several processes. 

Using the returned by the system call ACVA, the process can directly program any IO unit by 
means of writing into the control registers of such unit without using a driver and an operating 
system service. In particular, a process can define the access controlled addresses of DMA 
exchanges in the its own local or shared memory. Since any operation of writing into the register is 
always executed by some thread of the process with the RefPID identifier, the MIOMU can 
remember this identifier, thus noting that the IO unit acts on the orders of said process and hence 
all its DMA exchanges are performed identical as program threads requests. Further, the MIOMU 
channel dedicated for input-output device will form requests over access controlled virtual 
addresses in the same manner as in the method of program threads described above.


\chapter{ Synchronization over hardware-driven semaphores }

The preceding chapters 3,4 describe the two innovative decisions, which constitute the basis of the 
VThM architecture - the fine grain direct hardware multiprogramming and organization of inter-process communication by means of access controlled virtual addressing (ACVA). The current 
chapter presents the third basic decision - the direct hardware threads synchronization. The 
refinement of these three decisions is presented in chapter 7 in the context of the VThM processor 
reference model description.

The possibility of homogeneous interaction of threads belonging to different processes, 
implemented in the VThM architecture, makes it possible to improve the classical Dijkstra's 
\cite{Dijkstra68} approach of semaphore-based synchronization and to use it for the effective hardware 
synchronization of the GPPC. The proposed improvement is that the semaphores in the VThM 
architecture are implemented as active microarchitectural units, which are built into the MIOMU. 
These units, which are thereafter called the hardware-driven semaphores (HWDS), form a special 
microarchitectural independent activity units' pool, thereafter called the HWDS pool. 

In accordance with the fundamental work of Dijkstra \cite{Dijkstra68}, describing the essence of 
interaction of sequential processes, the correctness of each synchronization protocol should remain 
valid when the agents' relative running rates are changed within the range from the zero to the infinity. 
The requirement of the protocol efficiency is that the overheads of the synchronization should be 
minimal. Basically it boils down to the necessity to ensure that a thread waiting for an event 
associated with synchronization, does not occupy a processor. These requirements are fully 
implemented in the set of hardware features of the VThM architecture described below. An 
additional benefit of these features is that they free up the processor from switching threads 
between the waiting and activity states. These features at the ISA level are presented by the 
following six instructions - SemaphoreGet, SemaphoreFree, SemaphoreLock, SemaphoreUnlock, 
SemaphoreWait and SemaphorePass.

The SemaphoreGet instruction fetches a free HWDS from the pool, sets it into an initial state and 
returns its access controlled virtual address (ACVA). This address defines the semaphore as a part 
of context of the process, which is the owner of the thread issuing the SemaphoreGet instruction. 
Below this address will be used as the main operand in the remaining synchronization instructions. 
The SemaphoreGet instruction returns the empty result if there is no a free semaphore. 

The SemaphoreFree instruction makes the semaphore ready for reallocation. The HWDS pool, 
similarly with the pool of created threads, is implemented as virtual in the VThM architecture. 
Term virtual here means that HWDS pool is implemented as a multilevel facility similar with the 
described above VThM register files - the elements of pool can be swapped between levels 
purely by microarchitecture.

The HWDS has a register memory, which can be viewed as a set of described below structured 
variables, which are changeable only by means of microarchitectural features. The mutex variable 
of the semaphore is an analogue of the mutex variable in modern Unix-like operating systems 
\cite{SunOS}, \cite{Linux}. This variable contains the semaphore-owner field, which has an empty value if the 
critical interval guarded by the semaphore is free, or the identifier TID of the critical interval 
thread-owner.  The mutex variable also contains the pair of fields which have an empty value or 
define the FIFO threads queue, which is ordered by priorities and issuing times of the 
SemaphoreLock instruction described below, in a situation when a critical interval guarded by the 
semaphore is busy. The event variable is an analogue of the conditional variable in the mentioned above 
operating systems. It contains the pair of fields, which either have an empty value or define the 
queue of threads, which executed a SemaphoreWait instruction in the critical interval guarded by 
the semaphore,  and are ordered by priorities and issuing times. 

The semaphore also contains the counter variable, which obtains an initial value by performing the
SemaphoreLock ore SemaphoreWait instructions. Immediately after being set the counter 
variable starts to decrement with some frequency purely by means of hardware. Its turning into the zero 
before the issuing of SemaphoreUnlock or SemaphorePass instructions over the semaphore, forces  completion of 
SemaphoreLock and SemaphoreWait instructions with the corresponding completion code. 

The SemaphoreLock instruction provides an entry of a single thread into the critical interval guarded 
by the semaphore passed as the instruction's parameter. If such attempt is performed by an another 
threads before the 
first thread has left the critical interval by means of issuing the SemaphoreUnlock or SemaphorePass 
instructions, these threads will stand in the tail of waiting queue, which is identified by the mutex 
variable of the semaphore.

The SemaphoreUnlock instruction withdraws the issuing thread from the critical interval, thus 
allowing a new thread to enter the critical interval; this instruction is essentially the "close 
parenthesis" for the SemaphoreLock instruction.  

The SemaphoreWait instruction atomically performs an action of SemaphoreUnlock instruction 
and moves the issuing thread to the tail of waiting queue, which is identified by the semaphore 
event variable. The term "atomically" here and thereafter designates the sequence of actions, which 
are indivisible at the ISA level for threads cooperating via the same semaphore. 

The SemaphorePass instruction atomically performs the following actions:

- removes the instruction issuing thread from the critical interval;

- browses the  waiting queue related to the event field, and if this queue is not empty, enters 
the first thread from this queue into the critical interval;

- if  the queue related to the event field is empty, enters the first thread from the queue 
pointed by the mutex field into the critical interval;

- if both queues are empty, makes the critical interval free.

An introduction of the four synchronization instructions instead of the two proposed by Dijkstra \cite{Dijkstra68}
analogues of instructions SemaphoreLock and SemaphoreUnlock, is stipulated by the following 
considerations. The thread, having executed the instruction SemaphoreLock, can find the lack of 
data for processing. In this case, the thread must re-enter the critical interval after having been waited 
for some time. The number of such transitions is not known in advance. Therefore the real time system dilemma 
is appeared again - either to perform a polling frequently thus increasing overheads or perform a polling 
rarely thus increasing a probability of loss control. The implementation of the pair of 
additional commands SemaphoreWait and SemaphorePass in the ISA permits to enhance 
significantly the efficiency of synchronization in special cases of waiting for an event of appearing 
the first element of information. 

It should be noted that SemaphoreUnlock is the  special case  of SemaphorePass instruction. We suppose that it instruction is useful for simplest microprocessors in which sophisticated SemaphoreWait and SemaphorePass instructions may be omitted.

The c language source code shown on figure 1 explains the usage of synchronization instructions   
proposed above in order to program the classical supplier-consumer algorithm. The procedure 
VThMSyncDemo starts a pair of threads by means of the CreateThread system call; it also 
creates a semaphore for their interaction an also completes the task. It seems that the comments 
accompanying the c code illustrate the logic of the algorithm as well as the logic of used VThM 
ISA synchronization features in sufficient details.

\begin{figure}[h] 
\begin{center}
\scalebox{.5}{\includegraphics{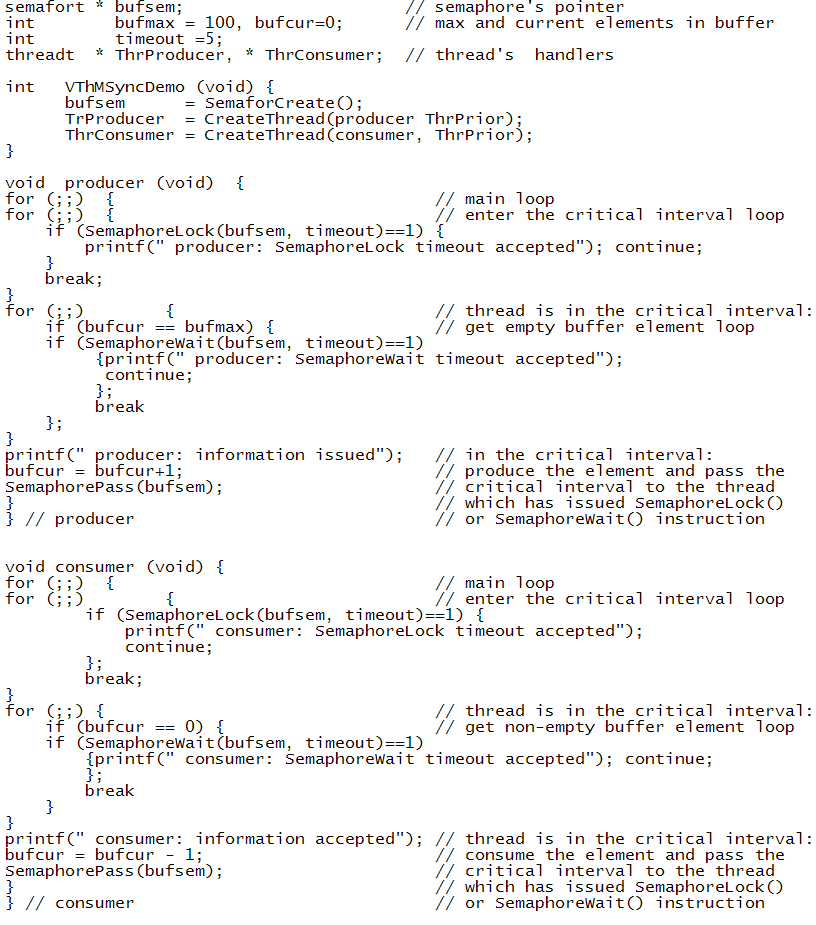}}
\caption{ Producer-consumer synchronization algorithm }
\end{center} 
\end{figure}

To sum up everything described in this chapter, it is important to note that the HWDS as the active 
microarchitectural level unit allows to eliminate the important bottleneck present in the modern 
day implementations of GPPC - the centralized scheduling and dispatching of a set of threads by 
means of an ISA code of an operating system kernel. Essentially the HWDS performs the function of local 
scheduling of the access to the critical interval associated with a semaphore and in cooperation with the 
microarchitectural sequencer unit described below in chapter 7, implements the local dispatching of the 
subset of threads, interacting through the semaphore. The HWDS while implementing hardware 
timing of waiting states associated with critical intervals, also significantly improves the 
centralized operating systems \cite{SunOS}, \cite{Linux} time service, making it per-semaphore distributed 
and purely microarchitecturally implemented. 


\chapter{  IO programming without interruptions  }

This chapter describes the innovative technique of IO programming based on the introduced above hardware driven semaphores (HWDS), which allows 
to eliminate the second, and seemingly the last, bottleneck in the existing GPPC implementations 
- the centralized support of the IO programming by means of an operating system ISA code, which uses 
interruptions. This technique is demonstrated on the basis of the simplified variant of supplier-consumer algorithm described on the figure 1 of previous chapter.

The threads discussed in this algorithm perform the processors' instructions and thus are 
essentially computational threads; the algorithms of their cooperation are completely 
symmetrical. The data exchange with external devices is performed by means of hardware threads, 
thereafter called the IO threads. The principal feature of the IO threads is their total 
subordination to the computational threads. This subordination means that in a correctly designed 
and properly functioning software-hardware environment nothing activity of IO devices can occur 
without the computational threads control. 

In more detail, after resetting a computer all IO units are reseted into the idle state. An exit of these 
units from this state occurs only when a program thread has written into a transfer control registers the following information: an IO operation code of the block to be transfered, its size and addresses  in the process-owner memory and in the space of  IO unit.

Having obtained this information the IO unit starts the 
execution of the IO thread, which performs the data transfer operating in the direct memory access 
mode (DMA) and informs about the completion of the transfer by issuing a processor interruption. 
In existing architectures this interruption is processed by means of an operating system kernel and 
device driver. The final result of this software and hardware handling of interruptions is the 
transfer of the IO related thread from the waiting state into the active state. There are three alternatives for the 
data exchange completion - the normal transfer of the data block, the transfer with errors and 
the completion upon time-out.

Using the principal feature of the VThM architecture, namely its ability to support effectively the 
execution of a virtual set of threads, it is possible to suggest the following simple and effective 
technique for organizing the data exchanges without interruptions as follows. So-called the dual 
computational thread is created for each IO unit channel with an independent hardware activity. This 
thread initiates and completes active phases of the affiliated IO thread for each data exchange. 

If a system provides a unique identification of IO unit, which has completed the transfer (for 
example, the protocol MSI of PCI Express bus), the dual computational thread and IO thread can 
only interact with each other. If a legacy interruptions technique is implemented via a fixed 
number of lines as in the PCI bus protocol and protocols of older buses, then for each interruption 
line the computational dual thread-multiplexer is created.  On one hand, this thread interacts with the 
interruption delivery logic and executes an interruption confirmation protocol, and on the other 
hand this thread interacts with a set of dual computational threads of all IO units, dedicated to the 
same interruption line.

The technique of semaphore-based synchronization described in the preceding chapter allows to
completely eliminate the program processing of interruptions by the kernel and the drivers of operating 
system. The basic idea is that an IO interruption processing hardware is supplemented by a special 
unit, thereafter referred to as the interruptions control unit (ICU). This unit organizes an interaction of IO 
devices with the dual thread by means of the access controlled virtual address (ACVA) of 
semaphore, which address is stored by the computational thread at the dedicated register of the ICU. In terms of the discussed above supplier-consumer algorithm, the 
simplest scheme of a purely hardware-based interruptions processing is as follows. 

Each interruption to be processed in accordance with this protocol is represented by the 
interruption control block (ICB) defined below. These blocks are cells of specialized units built in the 
MIOMU. They are analogues of the ICB and the hardware driven semaphores described in the previous chapter. These blocks are 
allocated and released by means of the GetIcb and FreeIcb instructions respectively. The role of the 
supplier of interruptions is performed by the ICB, and the role of the 
consumer of interruptions is performed  by the dual computational thread. The exchange buffer and its 
descriptive information boil down to the binary counter allocated in the ICB.

Shortly, the dual thread writes its own thread identifier, the zero counter value and the semaphore 
address, into the ICB block. In the simplest case of PCI interruption delivery protocol this 
information is INTA-INTD lines number. In the more sophisticated case of MSI interruption 
delivery protocol, this information comprises of an interruption related bus, unit and function 
numbers. 

At the moment of interruption arrival, the supplier issues the SemaphoreLock synchronization 
instruction into the MIOMU. This instruction uses the address of the interruption-dedicated 
semaphore stored in the ICB, as the first operand. Then after having entered the critical 
interval, the supplier normally discovers the zero value of the ICB counter, changes it to the non-zero 
value and issues the SemaphorePass instruction. By this moment the dual thread will certainly be 
in the state of waiting associated with the semaphore event as a result of execution of the 
SemaphoreWait instruction.  

Correspondingly, having found the non-zero value of the counter, which corresponds to the 
interruption processing cycle unfinished by the dual thread, the supplier will perform atomically 
the sequence of actions, which comprises of issuing the SemaphoreWait synchronization 
instruction into the MIOMU and switching the supplier thread from active state into waiting state.

After that the consumer, whose role is performed by the dual thread, is transferred into active state, 
following a completion of the SemaphoreWait instruction whose result indicates the arrival of the ICB related
event. Having analyzed the ICB counter and having normally found the non-zero value, the consumer 
rewrites the interruption related information into the ICB related program buffer, performs the interruption 
confirmation protocol, assigns the zero value to the ICB counter, and notifies the supplier about the 
release of the interruptions delivery logic by means of issuing the SemaphorePass instruction.

It should be noted that in modern microprocessors, in particular in the x86 family, the delivery of inter-processor interruptions is implemented uniformly with the delivery of interruptions from IO devices 
by means of the APIC unit. It allows to extend the IO interruptions processing techniques 
described above onto the inter-processors interactions and eliminate the interruption processing. 
It letting to introduce the new of a programming of  an IO and interprocessor communication, 
which in a steady state of control over a plurality of created threads and processes  is free of interruptions, operation system and drivers support. 


\chapter{ The VThM processor reference model }

This chapter presents the VThM processor reference model. In this model the main microarchitectural 
units and logic of their interaction at the minimal refinement level, necessary for precise 
illustration of all described above decisions in commonly used terms of the circuit engineering, are 
described. 

Figure 2 presents the general block diagram of processor in the VThM architecture. 
Functions of the described above conceptual entities - the process engine and thread engine - are 
implemented as the distributed protocol by means of a hardware units.

The VThM 
processor consists of a set of thread monitors, a set of domain executive clusters and a single 
combined memory and input-output management unit (MIOMU) which supports data transfers  
based on the access controlled virtual addresses (ACVA) introduced in the chapter 3. With the purpose 
of the effective support of all kinds of software debugging, the hardware debugging monitor is used. 

All 
devices interact through a processor network, which is the broadband packet switching network. The 
principal feature of this network is the hardware-implemented priority scheme, in which a multi 
channel router supports the transmission of all information elements between all other units in 
accordance with the priorities of the producing threads. 

\begin{figure}[h] 
\begin{center}
\scalebox{.4}{\includegraphics{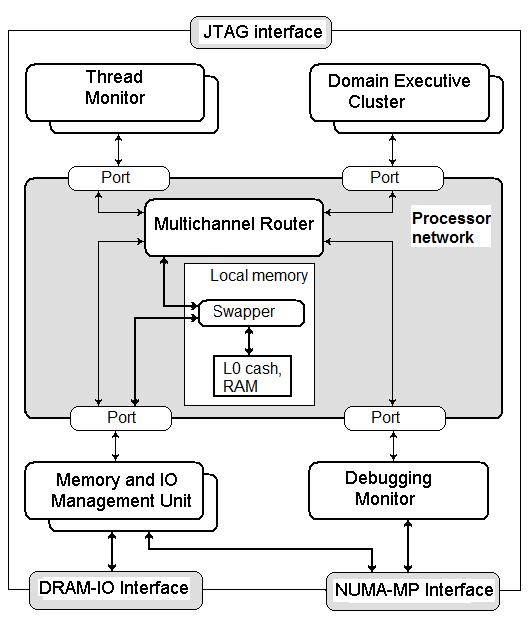}}
\caption{ VThM processor general block diagram }
\end{center} 
\end{figure}

As it will be explained in the next chapter, the VThM processor microarchitecture is essentially the 
dataflow machine, which is well suited for implementation in silicon using the processor-in-memory 
(PIM) technology. Therefore at the microarchitectural level all units, including the processor 
network, have a microarchitectural virtual memory. This memory is used for hardware swapping 
of microarchitectural information in accordance with its priority and the aging metric in circumstances of 
competition for levels of microarchitectural memory, processing units and transfer logic of the network. This 
memory is shown on all the presented below figures as a local memory, and is not mentioned 
further in the description of these devices. A swapper unit implements the interaction over 
exchange of information with connected to this memory devices as well as the movement of 
information between the levels inside the microarchitectural virtual memory. 

The VThM processor has the following external interfaces. The DRAM-IO interface of the 
MIOMU unit provides for an interaction with the main RAM and IO units. The NUMA-MP 
interface provides for the merging of the VThM processor into multiprocessor systems with the non-uniform memory access (NUMA). The JTAG interface is used as usual for the debugging 
interaction with the processor's hardware. It is also used as an interface of interaction of the 
debugging monitor with corresponding software, executable on the instrumental machine.

Figure 3 shows the thread monitor that consists of a sequencer and a transaction issuing unit. The 
sequencer consists of an operating registers block for storing the roots of thread descriptors and of 
a scheduler unit. The monitor provides an automatic creation of  threads after hardware reset of a 
computer and a programmed creation of a new thread while performing an ISA instruction in 
accordance with described in the chapter 5. 

\begin{figure}[h] 
\begin{center}
\scalebox{.4}{\includegraphics{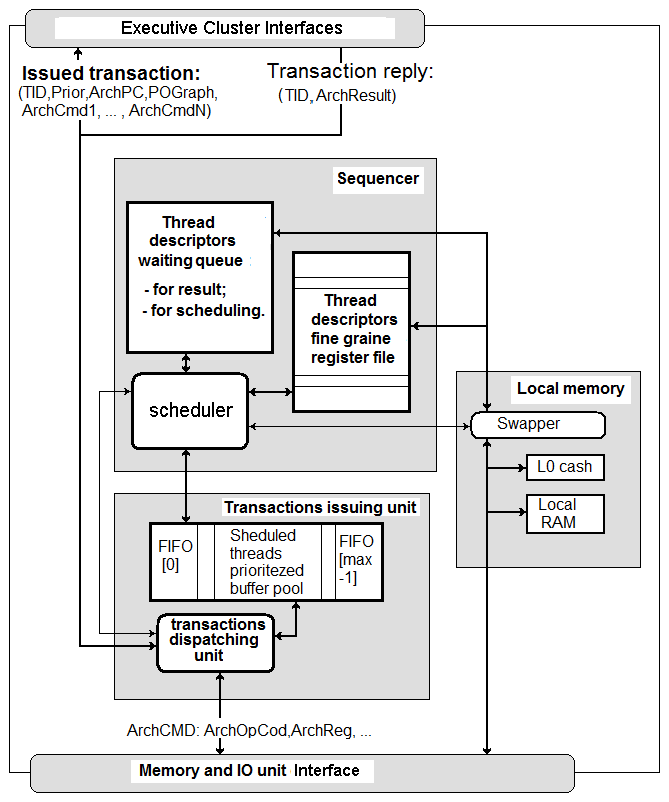}}
\caption{ Threads monitor block diagram }
\end{center} 
\end{figure}

Creation of a thread in the microarchitecture level boils down to the issuing a request for the 
establishment of a process descriptor into the MIOMU. If the positive acknowledgment is accepted, 
the monitor creates a root of the thread descriptor block in the sequencer's operating registers block. 
This root contains the minimum of information necessary for issuing a new transaction in accordance 
with the result of the completion of the previous transaction. An additional information, which is 
not included into the root, is stored in a distributed form in the different executive clusters and in 
the MIOMU. For example, ISA register files can be distributed as follows. The integer registers 
are placed in the cluster of integer arithmetic, and the registers for the floating point operand are 
placed in the floating-point arithmetic cluster. 

The information stored in the root essentially depends on ISA implemented by the monitor. At the 
very least, it contains the thread identifier TID, priority, status of privileges, counters of current 
and next instructions and the completion codes register for performing conditional branches.

The roots of the thread descriptors are placed in the operational registers blocks which are 
organized in waiting queues of two types. In the first queue the descriptors of treads, which wait 
for scheduling the performance of the next transaction, are stored. In the second queue the 
descriptors of the treads, which wait for completing the earlier issued transaction, are stored. The 
scheduling of the instruction for execution boils down to fetching an ISA code segment in 
accordance with the current value of instruction counter, which is required for formation of a 
transaction. The transaction represents a set of architectural instructions to be executed and  an 
information dependency graph, which specifies a partial order of execution of this set of 
instructions. This complexity in the VThM architecture is used to ensure speeding up the 
execution of high priority thread instructions in accordance with WLIW and super-scalar 
approaches.

An important feature of the VThM architecture is its ability to implement a fine grained dynamic 
physical resources allocation in accordance with actual demands of a code scheduled for 
execution. This is completely analogous to the dynamic allocation of virtual memory pages 
demands. For example, an interruption handler for the RISC architecture, well-coded in assembler 
language, can use only the global registers. Therefore hardware resources for other ISA resources 
will not been allocated for the interruption handler thread in the VThM microarchitecture.

The scheduler knows the ISA of instructions stream, and in accordance with the size of transaction 
and with the semantics of its instructions, the scheduler defines microarchitectural resources 
necessary for its performing. If another transaction accesses the microarchitectural resources that 
are not yet allocated, the scheduler issues the special transaction into the executive cluster or into the 
MIOMU for their allocation.

The scheduler stores the formed transaction into the prioritized queues buffer pool placed in the 
transaction issuing unit. This unit supplements the transaction with information, which on non-optimized conceptual level contains the following elements: 

- the thread identifier TID necessary for unique identification of the transaction and for its each  
instruction within the framework of a distributed protocol, performed by a set of the
microarchitectural units; 

- the architectural number of each instruction necessary for the precise localization in terms of the ISA 
instruction, which caused an abnormal situation; 

- the priority of the thread-owner, which determines an order of the thread's elements servicing in the all 
competition circumstances.

The transactions dispatching unit sends a transaction for the execution as well as receives its 
results upon the transaction completion. This unit interacts with the scheduler via the prioritized 
queues buffer pool placed in the operating registers of the transaction issuing unit. The result of the 
transaction completion is stored in the field known to the scheduler. The scheduler makes free 
all the operational registers blocks have been occupied by the completed transaction after adoption the 
transaction's completion result.

The use of transactions allows to localize an execution of an ISA code fragments in a separate 
domain executive clusters, thus significantly reducing the traffic in the VThM processor network. The good 
example in this respect is the implementation of all functions of a Unix like operating system 
string processing libraries as single transactions, which can be fully carried out by an integer 
arithmetic executive cluster. At present the whole processor is required to perform the functions of 
these libraries, written more than 30 years ago. At the same time because of the achievements in 
digital circuit engineering, significantly more complex USB  \cite{USB} and SATA \cite{SATA}  protocols are 
implemented by a very simple and inexpensive controllers. Obviously, the development of a string 
processing cluster or a specialized unit within the MIOMU should significantly improve the efficiency 
of the GPPC implementations. 

This example demonstrates the fact that the VThM architecture allows the very effective usage of 
both CISC and extended CISC architectures, which are now considered as erroneous. This 
efficiency would be achieved by means of distribution of parallel execution of a large number of 
architectural instructions streams. It seems that a further research in this direction could contribute 
towards significant acceleration of a computation of any nature on the basis of its transfer to the ISA 
implementation.

Figure 4 presents the block diagram of the domain executive cluster. The cluster consists of a 
sequencer and an asynchronous executive pipeline. The sequencer consists of a mapping unit, a 
register file of waiting queues and a scheduler. The executive pipeline consists of a register file 
(RF) of instructions, a register file of operands, a set of functional executive units (FEU) and a 
load-store unit (LSU). The register file of instructions contains a buffer pool of ready-to-run 
instructions queues, which are ordered by instructions' priority and arrival time.

\begin{figure}[h] 
\begin{center}
\scalebox{.4}{\includegraphics{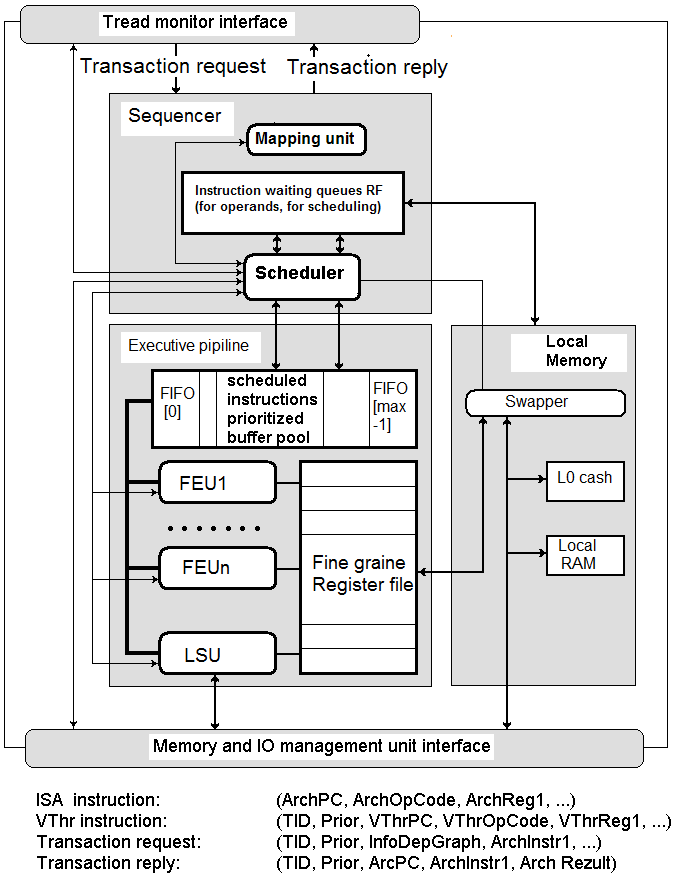}}
\caption{ Domain executive cluster block diagram }
\end{center} 
\end{figure}

The sequencer accepts the transactions-requests and rewrites them into the queue. These transactions 
contain the instructions to be performed in this cluster and the information dependencies graph 
describing a partial order of instructions execution. Among such instructions there could be instructions of 
local jumps within the transaction. 

An execution of jump instructions beyond the instruction block of a current transaction or an 
execution of the last instruction of the transaction leads to finalization of the transaction 
processing in the cluster and to returning of the result back to the thread monitor which issued the 
transaction. The threads monitors' pool and the executive clusters' pool together provide the two-level 
scheme of execution of instructions, which is efficient for the realization of heterogeneous 
processors with shared executive pipelines for legacy and native ISA. The execution of a local 
conditional jump instruction allows to aggregate the instructions and thereby to reduce the intensity of 
a traffic in the processor network.

The mapping unit performs the replacement of architectural registers addresses with the 
microarchitectural addresses of the cluster's register file allocated for the instruction's thread-owner, and transfers thus prepared instructions to the scheduler. Essentially the scheduler 
performs functions of distributed dispatching on the microarchitectural level; in present 
architectures these functions are performed by means of operating system ISA code. 

The principal difference is that instead of the known coarse grain program switching of threads 
between active and waiting states as an indivisible entity, a set of  the microarchitectural VThM
processor's schedulers performs the fine grain fragment by fragment switching and executing of the threads. 

Such a switching on the transactions performing level boils down to moving ready-to-run 
instructions (also called active instructions)  from the scheduler waiting queues into the executive pipeline 
input queues and moving in the reverse direction those instructions, which turn into the waiting state 
until their operands become ready or execution of the instruction become completed. Similarly the schedulers 
of described above thread monitors work at the threads performing level - the threads correspond to 
the transactions, and the transactions correspond to the instructions. The generalized effect of such 
a fine grain performance in the VThM architecture is described above in the chapters 2 and 3.

The scheduler uses the information dependencies graph, contained in the transaction, for fetching 
the next ready-to-run instruction. The executive pipeline corrects this graph after execution of 
each instruction. Following completion of the transaction, the resources, which have been used for 
storing the instructions and the graph, become free. In case of the appearance of a higher priority 
transaction, the scheduler can force out an instruction of a lower priority transaction from the input 
queue of the executive pipeline and place it back into the waiting queue. Instructions and elements 
of information dependency graph, which remain inactive for a long time, can be forced out into 
the lower level of the local microarchitectural virtual memory of the cluster. The operational 
registers block allocated to any thread can be forced out after the last instruction that used this 
block, has been forced out.

The load-store unit (LSU) provides the data exchange between microarchitectural operating 
registers and the MIOMU. Usual memory access instructions, in particular those in systems with 
non-uniform memory access (NUMA), may be very long lasting. Essentially the synchronization 
instructions are the specific memory access instructions, which force the hardware driven semaphore 
(HWDS) logic to perform the actions described in the chapter 5 and  to move the instruction into 
the state of waiting for the reply from semaphore logic. Such waiting states are the states of 
algorithmic latency described in chapters 1 and 2, which may be very long lasting. This can force the
downloading of all distributed thread representation elements from the MIOMU, the executive 
clusters and the thread monitors into the lower levels of microarchitectural virtual memory.  

At some moment the HWDS will post the synchronization instruction completion code into the 
load-store unit. This event will induce the predefined by instructions sequence of uploading of the
distributed threads representation elements into the highest level of memory, e.g., into the 
operational registers. These uploading will force the execution of the next steps of the thread running 
by the set of microarchitectural schedulers and executing pipes.

Figure 5 presents the block-diagram of a memory and IO management unit (MIOMU). This unit 
consists of an access validation unit, a ACVA translation unit, a synchronization unit, and a 
physical memory and IO control unit. The later consists of a routing unit, a block processing unit 
and an interruption unit. 

\begin{figure}[h] 
\begin{center}
\scalebox{.4}{\includegraphics{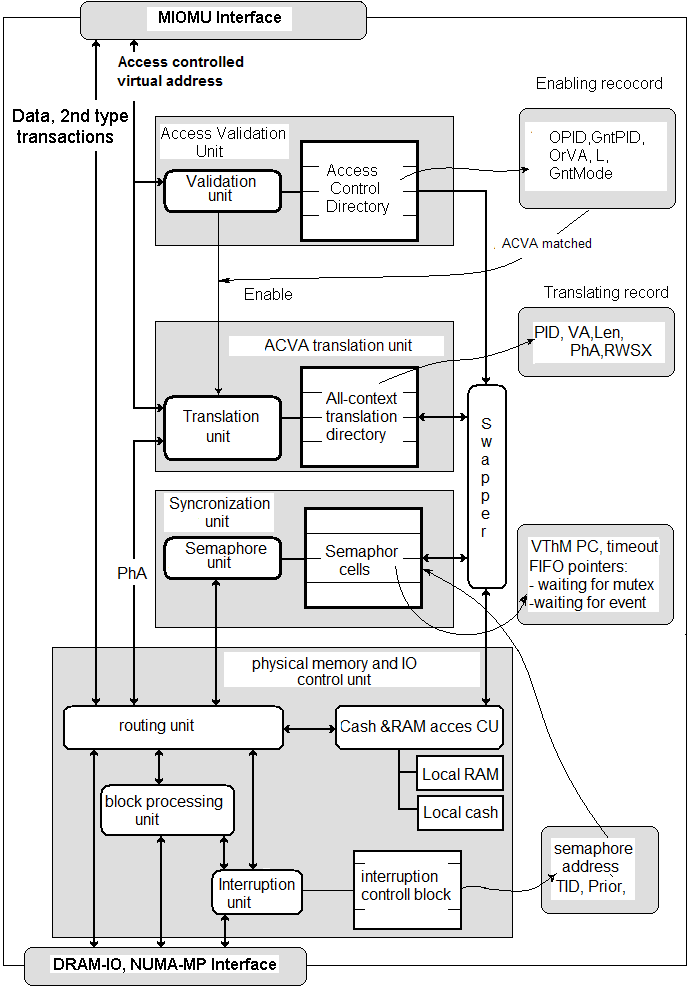}}
\caption{ Memory and input-output management unit }
\end{center} 
\end{figure}

The MIOMU performs two types of transactions. Transactions of the first type present a single 
access of the threads monitors or executive clusters via of access controlled virtual address 
introduced in the chapter 4. Such address arrives simultaneously into the validation and the 
translation units. The validation unit matches associatively this address with the contents of its 
access control directory and on the basis of this matching it issues the signal of access permission 
or prohibition into the translation unit. The access matching procedure have been detalized in the chapter 4.

The translation unit performs the conversion of the access controlled virtual address into the 
physical memory address or the IO unit registers address. If the validation unit issued the signal of the
access permission, the translation unit performs the corresponding access operation or informs about 
the abnormal completion of the operation. The functions of the translation unit of the VThM processor 
are similar to the functions of the existing processors' virtual memory management units 
(VMMU) but unlike the latter they use the advanced translation directory in order to provide 
simultaneous support of contexts of all created processes. The validation unit is one of the most 
important innovative elements of the VThM architecture. It should be noted that all permitted 
accesses to the present virtual memory pages, processor and IO units registers are performed very fast by 
means of the direct hardware implementation. A program support is required only for rare accesses to 
the pages of virtual memory that have been unloaded from RAM, and for unauthorized accesses, 
resulting in an abnormal termination of a thread. 

The synchronization unit consists of a multi-channel HWDS engine and a semaphores register file. 
The HWDS engine channels perform the synchronization instructions discussed in chapter 5. The 
register file consists of semaphore cells, which contain the structural variables, semantic of which 
is described above in chapter 5. The thread descriptors' queues are placed in the MIOMU local 
virtual memory unit, and their exchange with the latter is supported by the swapper unit.

The transactions of the second type have the structure which is fully analogous to the structure of the 
transactions described above, which are coming into the executive clusters, with the only 
difference that they contain specific instructions, thereafter referred to as the block processing 
instructions. Such transactions are performed by the block processing unit, which essentially is an 
improved version of the existing direct memory access units. This unit allows to implement the data 
block transfer operations of memory-memory, memory-IO and IO-IO types at the 
microarchitectural level. In the existing architectures the functionality of these transactions is 
performed by means of software and takes a long time of a processor integer unit work, during 
which floating point units remain idle. Moreover, the encapsulation of the block processing in the 
MIOMU significantly reduces the processor's network traffic. As a very important and natural 
function of the block-processing unit it is logical to suggest the microarchitectural implementation 
of the virtual memory pages swapping.

The routing unit provides the redirection of the requests by a physical address into the appropriate 
physical unit - the executive cluster, a memory bank or the IO control unit. The physical memory 
control unit provides the data exchange with the memory banks. 

In addition to the known functions of the IO handling, the physical memory and IO control unit of 
the VThM processor performs the discussed in the chapter 6 protocol of transformation of interruptions 
into the synchronization operations using the hardware driven semaphores. These additional functions are 
implemented by a multi-channel active unit thereafter called the interruption unit. The 
interruption unit interacts with the dual threads using the interruption descriptors pool, each element of 
which represents the interruption allowed to be handled.  The dual threads activate these descriptors 
similarly to the semaphores described in chapter 5. These threads write the TID of IO thread, its 
priority, the address of the semaphore and the zero value of counter into the appropriate fields of the
interruption descriptor. Writing this value is the activation signal, which forces the channel of 
interruption unit, affiliated with the interruption descriptor, to perform the protocol of the IO 
processing without program interruptions handling as described in the previous chapter.


\chapter{ Conclusion }

Concluding our description of the VThM architecture, we summarize the main decisions, present 
pragmatic factors that define its viability and a possibility of a rapid spread in practical 
implementations and also define its place in modern taxonomies.

The main positive feature of the proposed architecture is the high level of generalized latency 
tolerance defined in Chapter 1.  This tolerance neutralizes the all factors noted by Sterling 
\cite{Sterling05}, which reduce the real speed of a computations as follows.  Latency and overheads are 
minimized by the fine grain and deeply virtualized representation of the ISA in the microarchitecture. 
The starvation is precluded by the creation of a virtual set of active threads allowing to create the 
intensive flow of instructions sufficient to saturate all of the processing devices. The competition 
is organized properly by means of supporting the global priority processing discipline at the 
hardware level. This discipline ensures that parallel threads running without the priority inversion. 

The very important positive side effect of the proposed architecture is the considerable simplification 
of programming due to abandonment of the concept of interruptions in organization and 
programming of IO in operating systems.

The major pragmatic features, which determine viability and a rapid spread of VThM 
architecture, is ensuring a continuity in the relation to the existing software and the ability to create 
in bootstrap mode the new, more efficient, system and application software. These features can be 
achieved through the implementation of heterogeneous processors, which would support the existing 
legacy ISA with the high degree of precision and would provide to realise their native VThM ISA. 
Operating systems for a legacy ISA can be executed on virtual machines that are implemented in the 
VThM architecture fully by the means of hardware. New operating systems and applications 
software for the native VThM architecture should provide a better performance in the case of
reprogramming of old tasks.

The important advantage of performing legacy applications under control of legacy operating 
systems at the heterogeneous processors is the possibility of direct use of the virtual-threading 
features in the important and widely used applications such as database servers and search 
engines. These features include the hardware synchronization of threads and the direct IO programming without 
participation of the legacy operating system, which can simply be not aware about the existence of 
these features. In the legacy ISA they can be represented as a special IO unit, which can be 
considered as a virtual-threaded co-processor. This unit will transfer the transaction which was formed by 
means of legacy ISA program, will send it into the necessary executive cluster and will return the result 
back to the program. The same mechanism can be used to an information exchange between 
applications on different operating systems by means of the access controlled addresses introduced 
in chapter 4. The ability to work effectively and independently for a plurality of operating system 
on a single VThM processor simultaneously is provided by the hardware implementation of well-known concept of virtual machines.

It seems that the concept of heterogeneous processors based on the VThM architecture, where 
different ISA programs can effectively communicate through the shared memory, should cause a 
review of existing assessments of a number of architectural solutions as erroneous. For example, 
the implementation in a powerful microprocessor, dedicated for network information retrieval 
systems servers (Web queries) or a large database system, combination of the RISC and CISC 
(and even Extended CISC) ISA can lead to the following. An interaction with the 
network in part of receiving requests and issuing responses can be supported by RISC 
applications. Processing of requests related to fetching information from large databases will be 
performed by extended CISC applications where relational database model operations \cite{Codd69} 
will supported on the ISA level. In advanced implementations of such systems the functions of 
artificial intelligence - e.g., identification of semantic queries using knowledge database or 
translation from one language to another - can be additionally implemented. In these systems 
support of specialized languages such as lisp \cite{Lisp58}, refal \cite{Refal66} and prolog \cite{Prolog72} is 
justified. 

It seems that VThM architecture makes need to review a large number of decisions in processors 
architecture, which now are considered as erroneous.
The most important in this respect, proprietary the VThM architecture is the significant improvement 
of MTA-architecture, which can eliminate virtually all its shortcomings.

Finally, let's define the place of the VThM architecture in the existing taxonomies. With respect 
to the Flynn's classification of parallel processors \cite{Flynn}, it is naturally to consider the VThM 
architecture as an extension of MIMD architecture and to call it VIVD architecture - the machine 
with Virtual Instructions and Virtual Data streams. 

The VThM architecture extends the existing classification of processors with respect to the hardware 
virtualization and makes this virtualization two-dimensional. Indeed, in respect of memory 
virtualization on the level of ISA existing processors are divided into two classes - with 
the physical memory (PhMem) and with the virtual memory (VMem). It should be noted that we 
understand the virtual memory as the hardware features at the ISA level, which support context 
protection domains as well as translation of a virtual address into a memory or a register physical 
address. The cashing features on the microarchitectural level are considered as irrelevant to 
virtual memory. 

Processors of PhMem and VMem classes can support two classes of threading. First class we 
propose to call as the processors with the physical threading, which includes the known single-, hyper- and 
multi-threaded ones. Second class we propose to call as the processors with virtual-threading, which we 
introduce in this paper. We use fore these two classes abbreviation PTh and VTh respectively. 
Thus with respect to the virtualization, similar to the Flynn's classification in relation to parallelism, 
we  propose to define the four classes of processors:

PhMem-PTh - the classical von Neumann machine, according to which early computers were 
implemented.

VMem-PTh - the von Neumann machine with the virtual memory and the physical threading, which 
began to crowd out early PhMem-PTh machines in the mid-60s and has been used in the majority 
of modern processors.

PhMem-VTh - the machine with the physical memory and the virtual-threading at the ISA level, 
based on the virtual memory of the microarchitecture. It seems that this new architecture will be the most 
suitable for extremely hard real-time systems.

VMem-VTh - the two-dimensional virtualized machine with the virtual memory and virtual-threading at the 
ISA level. It seems that this new advanced architecture will replace the existing most widely used 
VMem-PTh architecture in a powerful general purpose microprocessors and will be the 
architecture-candidate for implementation of the general purpose supercomputer discussed in 
chapter 1. 

In respect of whether to classify the virtual-threading architecture as dataflow or von Neumann 
machine, we consider the following. At the ISA-level the virtual-threaded architecture supports a
simultaneous work of many traditional von Neumann machines, also referred to as the stored-program 
machines. These machines, or more accurately, the virtual processors that represent them, are 
stored and processed as a usual data at the virtual-threaded microarchitecture. Therefore the 
virtual-threading architecture can be seen as an improvement of the von Neumann machine 
architecture and it is natural to call it the stored-processors machine.

Further it is not difficult to see that at the level of microarchitecture the virtual-threaded  machine 
actually implements the dataflow machine. In essence, the data to be processed by this dataflow machine is the
fine grain distributed representation of processes and threads which are being stored in the 
multilevel virtual memory. The work of such a dataflow machine provides running of a virtual set 
of von Neumann machines. Those elements of ISA image of the von Neumann machines, which are 
required for an instruction performing, are fetched by means of hardware into the level of operating 
register and pushed into the opposite direction if the von Neumann machine remains inactive for a 
long period of time.  

Presumably this duality of the virtual-threaded machine, which is reflected in the combination 
of the dataflow machine at the microarchitectural level and the von Neumann machines at the ISA 
level, is natural for the effective support of the general purpose parallel computing. Because of the 
sequential nature of human thinking, programmers are positively disposed to programming the 
von Neumann machines, and negatively disposed to programming the dataflow machine, which 
resulted in very narrow spread of the latter. This duality of the virtual-threaded architecture puts 
a man and a machine in their natural places - programmers will write programs for this architecture just as for a usual von Neumann machines, whereas running of a plurality of these machines 
automatically generates a dataflow program which will be performed by means of virtual-threaded processor dataflow microarchitecture.


\backmatter

\end{document}